\documentstyle[aps,prl,twocolumn,psfig]{revtex}

\begin{document}
\draft                       


\title{Structure and dynamics of a model glass:\\ influence of 
       long-range forces}

\author{P.~Jund, M.~Rarivomanantsoa and R.~Jullien}

\address{Laboratoire des Verres - Universit\'e Montpellier 2\\
         Place E. Bataillon Case 069, 34095 Montpellier France}


\maketitle


\begin{abstract}
We vary the amplitude of the long-range Coulomb forces within a classical 
potential describing a model silica glass and study the consequences 
on the structure and dynamics of the glass, via molecular
dynamics simulations. This model allows us to follow the variation 
of specific features such as the First Sharp Diffraction Peak and the Boson
Peak in a system going continuously from a fragile (no Coulomb forces) to 
a strong (with Coulomb forces) glass. In particular we show that the 
characteristic features of a strong glass (existence of medium range order, 
bell-shaped ring size distribution, sharp Boson peak) appear as soon as 
tetrahedral units are formed.
\end{abstract}



\narrowtext

\section{Introduction}

The connection between the structure and the dynamics of a given system is 
of particular interest in glasses since the structural entities at the origin
of some specific vibrational features like the Boson Peak (BP) are not 
yet well identified. For instance in silica, the archetype of network forming
glasses, the existence of the BP has been connected alternatively to the 
presence of SiO$_4$ tetrahedra \cite{buch}, "domains" \cite{kivel} or voids 
\cite{elliott}. In molecular dynamics (MD) simulations the most important
ingredient is the force field used to mimic the interparticle interactions: the
aim is to reproduce as closely as possible the experimental vibrational data 
and then, by looking at the atomic positions, to give an interpretation in
terms of structure. In that sense the so-called "BKS" potential \cite{bks} 
has been widely used to describe silica and the results obtained on the 
structure \cite{struc,struc2}, the dynamical \cite{taras} and the thermal 
properties \cite{ct} compare well with experiments. This pairwise classical 
potential is the combination of a short-range term and the long-range 
Coulomb interaction (CI) which is obviously the most important ingredient of
the force field \cite{coulomb}. One remarkable property of this potential
is that CI forces alone are sufficient to reproduce the tetrahedral covalent
bonding network of the silica atoms without the addition of extra terms such as
angular terms. Our idea in this paper is to vary the 
{\em amplitude} of the CI (this is not strictly equivalent to varying 
the range) in order to study the variation of the structure and the 
dynamics of what can be seen as a toy silica glass. With this procedure
we can study the properties of a system going from a strong network glass
(the silica glass described by the usual BKS potential) to a fragile 
two-component soft sphere glass when the CI is turned off. The question we
would like to address is: how do the structure and the 
dynamics change along this path ? The aim is to study the connection between 
structural features like the First Sharp Diffraction Peak (FSDP) or the 
existence of a tetrahedral network and the BP, this connection being the topic
of contradictory interpretations in the literature \cite{elliott,soko,claire}.

\section{Simulations}

We performed molecular dynamics simulations for microcanonical systems
containing 216 silicon and 432 oxygen atoms confined in a cubic box
of edge length $L = 21.48$ \AA, which corresponds to a mass density of
$\approx 2.18$g/cm$^3$ very close to the experimental value of $2.2$g/cm$^3$. 
Periodic boundary conditions were used to limit surface effects. The particles
interact via the BKS potential whose standard functional form is,  
for two particles $i$ and $j$ (which can be either Si or O):
\begin{equation}
U(r_{ij}) = \frac{q_iq_je^2}{r_{ij}} + A_{ij}\exp(-B_{ij}r_{ij})
            - \frac{C_{ij}}{r_{ij}^6}
\end{equation}
where $r_{ij}$ is the interparticle distance, $e$ the charge of an electron
and the parameters $A_{ij}$, $B_{ij}$ and $C_{ij}$ are fixed as follows:
$A_{SiO}=18003.7572$ and $A_{OO}=1388.773$ eV;$B_{SiO}=4.87318$ and 
$B_{OO}=2.76$ \AA$^{-1}$; $C_{SiO}=133.5381$ and $C_{OO}=175.0$ eV\AA$^6$.
The partial charges used to compute the standard CI are given by 
$q_{Si} = 2.4$ and $q_{O} = -1.2$. This original form contains an unphysical 
divergence at very short distances which we have corrected by adding a short 
range repulsive term ($\sim 1/r_{ij}^{40}$).\\
In order to vary the strength of the CI (which is calculated using the Ewald
summation method) we simply multiply the partial charges 
by a parameter $\lambda$ which is fixed at the beginning of each simulation:
\begin{equation}
q_i \longrightarrow \lambda q_i
\end{equation}
A simulation starts from a liquid configuration at 7000K which is 
equilibrated during 50000 time steps (35 ps). This system is then cooled to 
zero temperature at a quench rate of $2.3\times10^{14}$ K/s which is obtained
by removing the corresponding amount of energy from the total energy of 
the system at each iteration. The aim here is not to obtain fine details
of the structure of real silica which explains the use of a rather fast
quenching rate compared to the rates that can nowadays be achieved. In any 
case computer quenches are still far away from those used in experiments.
The quenched glass samples are finally
relaxed in the microcanonical ensemble during 100000 time steps (70 ps)
after which the temperature has reached $\approx$ 2K.
We then average the structural informations over the last 20000 time steps
to avoid transient configurations and we diagonalize the dynamical matrix
corresponding to the final configuration to obtain the vibrational density
of states.\\
This procedure has been repeated for eleven values of $\lambda$ going from
$\lambda=0.1$ to $\lambda=1.1$. The ideal "$\lambda=0$" system would consist
of a soft-sphere like arrangement for the oxygen atoms (similar to the systems
we have studied earlier \cite{softsphere}) together with a random arrangement 
of the silicon atoms in the interstices of the oxygen structure. 
Because of the weak attractive O-O interaction present in the short-range part 
of the BKS potential, a small Coulomb interaction is necessary in order to 
stabilize this system: we found that a value of $\lambda$ around 0.065 is 
sufficient which explains why we started our investigation at $\lambda=0.1$. 
For this lowest $\lambda$ value we have checked on the radial pair distribution
functions as well as on other structural characteristics that the structure 
was indeed very close to the ideal arrangement described above.
On the other hand, strong 
values of $\lambda$ lead to an artificial local densification of the samples 
which are non homogeneous (holes start to form in the structure): for this 
reason the maximum value of $\lambda$ is limited to $1.1$. Of course the 
standard BKS results are recovered for $\lambda=1$.

\section{Results and discussion}

In order to examine the ``quality'' of the tetrahedral network we first
study the tetrahedral O-Si-O angle $\theta$ which should be ideally equal 
to $109.47^\circ$ in a perfect tetrahedron as well as the Si-O-Si angle 
$\phi$ which measures the relative position and orientation of 
two neighboring SiO$_4$ tetrahedra. To numerically determine the bond angle 
O-Si-O (resp. Si-O-Si), we determine for each Si (resp. O) atom the two 
nearest O (resp. Si) atoms and we calculate the angle between the two 
corresponding segments Si-O (resp. O-Si), the result being averaged over 
all the Si (resp. O) atoms. The variation of $\theta$ and $\phi$ as a function
of $\lambda$ is plotted in figure 1. For $\lambda \leq 0.5$ it is obvious that
the SiO$_4$ tetrahedra are not formed and that the network is not connected: 
the system remains analogous to a soft sphere glass up to a threshold value 
of about 0.6. This can also be seen in 
the inset of figure 1 where the Si coordination number is plotted versus 
$\lambda$. This quantity has been calculated by the integration of the
Si-O radial pair distribution functions up to a cut-off distance of 2.3 \AA\
which corresponds to the first minimum for all the values of $\lambda$.
This coordination number is greater than 5.5 for $\lambda=0.1$ and is close to 
4 for $\lambda \approx 0.6-0.7$. Nevertheless for $\lambda \geq 0.5$ $\theta$, 
the tetrahedral angle, is almost constant and 
close to the angle in a perfect tetrahedron (indicated by the black arrow) 
while $\phi$, the angle between the tetrahedra, is continuously increasing
with $\lambda$. This indicates that the CI has a direct influence on the
medium-range order and once the tetrahedra (the building blocks) are created,
they form a connected network whose characteristics depend on the strength of 
the CI.
In agreement with a previous study \cite{struc2} the mean Si-O-Si angle is
close to $150^\circ$ for $\lambda=1.0$ (standard BKS description) slightly
larger than the experimental value of $\phi$ ($144^\circ$ \cite{phi} 
indicated by the hollow arrow). The calculated value closest to $144^\circ$
corresponds to $\lambda=0.7$ but this value is not sufficient to recover 
other structural features of vitreous silica as we will see below, 
therefore this is not an argument to modify the standard BKS charges.\\
A second way to analyze the connectivity of the network is to monitor the 
distribution of $n$-fold rings: an $n$-fold ring is defined as the 
shortest closed path of alternating Si-O bonds. Therefore an $n$-fold ring 
consists of $2n$ alternating Si-O bonds. The ring distribution as a function 
of ring size is plotted in figure 2 for 3 values of $\lambda$: 0.2, 0.5 and 1.
For $\lambda=0.2$ the ring distribution is typical of an open structure
with a majority of small rings (the estimated number of large rings should 
be taken with a pinch of salt since finite size effects due to the limited length of the simulation 
box bias the results). For $\lambda=0.5$ the distribution becomes
peaked for rings with a specific size (5 and 6) which extend over distances
greater than the inter-atomic distances: molecules are formed in the
system. Finally for $\lambda=1.0$ we find the usual distribution of rings,
peaked at sixfold rings and almost symmetric around $n=6$ \cite{rings}, 
typical of the silica network.\\
A complementary way of analyzing the structure is to compute the static
structure factor S(q) as a function of $\lambda$. We used the standard way of 
calculating S(q) \cite{epjb} and focused our attention on a common feature
in glasses: the First Sharp Diffraction Peak. This pre-peak is related to
medium-range structures in the glass and its origin has been the subject
of vigorous debate over a number of years. In figure 3 we represent the
amplitude of the FSDP as a function of $\lambda$ and in the inset we show
the shape of S(q) for $\lambda$ = 0.2, 0.5 and 1 in the vicinity of the 
FSDP, together with the experimental curve \cite{sf}. With decreasing 
values of $\lambda$ the amplitude of the FSDP decreases and one can say
that for $\lambda \leq 0.4$ this feature does not exist anymore in S(q).
Indeed the residual value of the amplitude for small values of $\lambda$ is
not significant and is an artifact of the Fourier transform as can be 
seen in the inset. Together with the results shown in figure 1, this
indicates that the FSDP appears when the tetrahedra ({\em already formed})
arrange themselves into a connected network.\\
All this structural information shows that below a certain amplitude of
the Coulomb interactions the structure of our toy silica glass has lost
all the medium or extended range features and is similar to a soft-sphere
glass. From this study it appears that the transition value of
$\lambda$ between a fragile and a network glass is around 0.5. 
Of course
the only proper way to describe the covalent bonding network of silica
atoms would be to use quantum mechanical calculations and therefore
no precise physical meaning should be attached to this threshold value
$\lambda =0.5$. This value
is linked to the specific form of the BKS potential which
does not contain 3-body terms and in which the Coulomb interactions are
sufficient to build the SiO$_4$ tetrahedra. Nevertheless the good quality of this ``simple'' potential has been demonstrated in a recent study which has 
shown that SiO$_2$ glass samples generated with the BKS
potential are good starting configurations for  {\it ab initio} 
quantum mechanical molecular dynamics calculations\cite{epjb}. 
One interesting consequence of our results is that one should be
cautious when screened Coulomb interactions are used to simulate real 
silica glasses. Indeed screened interactions are in a sense similar to
``weak'' interactions like the ones we have simulated here with the small
values of $\lambda$ and mistreat the sum in reciprocal space which is
made in the Ewald summation method. We show here that the screening should
not be too strong in order to simulate a realistic silica glass.\\

In parallel with the study of the structure we have calculated the vibrational
spectrum as a function of $\lambda$. During the diagonalization of the 
dynamical matrix we did not calculate the eigenvectors (for computer time 
reasons) but used the eigenvalues (always positive) to compute the 
vibrational density. We
were of course interested in the Boson Peak which appears to be a 
characteristic of glasses even though it can also exist in some crystals 
\cite{kivel}. This peak, seen in Raman and inelastic neutron scattering 
spectra, reflects an enhancement in the density of states compared to the 
Debye distribution and is located at frequencies around $\nu_B \simeq 1.5$
THz in silica. 
When varying $\lambda$, the energy (and the pressure since we consider a 
microcanonical ensemble) of our system varies appreciably
and this is reflected in the variation of the largest frequency of the 
whole vibrational spectrum, $\nu_{\rm max}$, which increases from 
$\nu_{\rm max}\approx$ 20 THz for $\lambda=0.1$ to 
$\nu_{\rm max}\approx$ 40 THz for
$\lambda=1.0$. As expected, also the shape of the spectrum varies
with $\lambda$: for $\lambda$ small we have
basically one band since the oxygen and silicon atoms are almost identical
whereas for large values of $\lambda$ we observe a clear separation between
an acoustic and an optical band typical of the formation of well defined Si-O
bonds. But the Boson peak is in all cases clearly present.
However due to the $\lambda$ dependence of the whole spectrum, it is not 
legitimate to compare the BPs directly and therefore
we have first determined the normalized density of states
$g(f)$ using reduced frequencies $f=\nu/\nu_{\rm max}$.                          
The normalization is such that $\int_0^1 g(f) df = 1$.
Then we have used a standard way to extract the BP from the vibrational spectrum
which is to plot $g(f)/f^2$ because in the
Debye approximation $g(f) \approx f^2$ at low frequency. This curve has then
been fitted by a ``generalized Lorentzian'' \cite{ldv}:
\begin{equation}
I=I_0 f^n \cdot
 \frac{1}{[f^2+f_0^2]^m }   
   \label{eq:lorgen}
\end{equation}
which has been used to extract the position of the BP and the 
corresponding height
in a straightforward manner. In figure 4 we show the $g(f)/f^2$ data 
together with the fitting curves for three typical values of $\lambda$ and 
in the inset
we report the position and the intensity of the BP as a function of $\lambda$.
As expected, due to the normalization of the $g(f)$ curves,
the variations of the position and the intensity have opposite trends.
Apart from a first decrease of the intensity
between $\lambda=0.1$ and $\lambda=0.2$ that
we cannot interpret simply (at very small values of $\lambda$, there are 
some artifacts due to the form of the potential as indicated above), one
observes a net sharpening of the BP when $\lambda$ becomes larger than $0.5$.
This seems to indicate that a strong BP is linked to the existence of 
tetrahedral units or to the presence of an FSDP and therefore supports 
previous interpretations for the origin of the BP in silica \cite{buch}.  
Moreover the global variation of the BP characteristics with $\lambda$ 
confirms the recent experimental observation that the BP increases when 
going from fragile to strong glasses \cite{soko2}.
Of course these results, based on a comparison between reduced spectra,
 should be taken with care: in addition to the increase of the bandwidth 
when increasing $\lambda$ there is also a deformation of  the whole            
spectrum which is difficult to take into account simply. Finally, if 
some significance can be given to the apparent decrease of the BP intensity 
with increasing $\lambda$ for $\lambda > 0.7$, 
one could conclude that, once the tetrahedral network is formed, the increase of the medium range order leads to a decrease of the BP intensity.
Of course, such a statement needs to be validated by further investigations.


\section{Conclusion}

In conclusion, we have performed classical molecular dynamics simulations 
in a toy silica glass model in which we have changed the intensity of the 
long-range Coulomb forces. With this procedure it is possible to analyze the 
structure and the dynamics of a fragile glass (without Coulomb interaction) 
and a strong glass (with Coulomb interaction) within the same 
system.
We have shown that the formation of a tetrahedral network, the existence
of a First Sharp Diffraction Peak in the static structure factor,           
a characteristic peaked ring size distribution, as well as a relatively
sharp Boson peak, are  correlated to the strength of the Coulomb forces: all 
these features are absent in the fragile glasses and appear once the Coulomb 
forces exceed a potential-dependent threshold.  
Of course our study is limited to a very specific model and other ways of 
creating locally ordered units could have been explored. Nevertheless it 
sheds light on what is really happening in a system which can 
 be considered as either a fragile or a strong glass. By tuning
one single parameter (which therefore appears to be the predominant 
ingredient in the simulation of glasses) one is able to connect structural
features like the FSDP to vibrational specificities like the Boson peak and
give answers to questions previously raised by experimental studies.

Part of the numerical calculations were done at the computer center CINES
(Centre Informatique National de l'Enseignement Superieur) in Montpellier.


%
\newpage
\vspace*{1cm}
\begin{figure}
\psfig{file=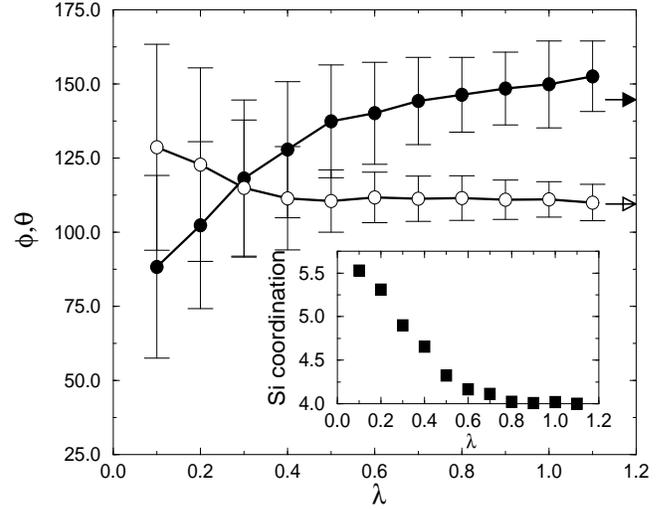,width=8.5cm,height=7cm}
\caption{
(a) Variation of $\theta$ (O-Si-O angle : {\Large $\circ$} ) and $\phi$ (Si-O-Si angle
 : {\Large $\bullet$}) as a function of $\lambda$. The filled and hollow 
arrows indicate the perfect tetrahedral value of $\theta$ and the experimental
value  of $\phi$ respectively. In inset the Si coordination number is shown as
a  function of $\lambda$.
}
\label{Fig1}
\end{figure}
\vspace*{1cm}
\begin{figure}
\psfig{file=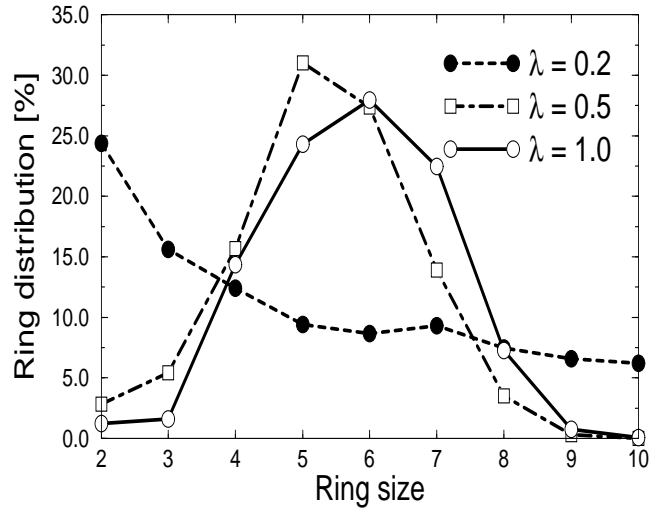,width=8.5cm,height=7cm}
\caption{ Ring size distribution for $\lambda$ = 0.2, 0.5 and 1.
}
\label{Fig2}
\end{figure}

\begin{figure}
\psfig{file=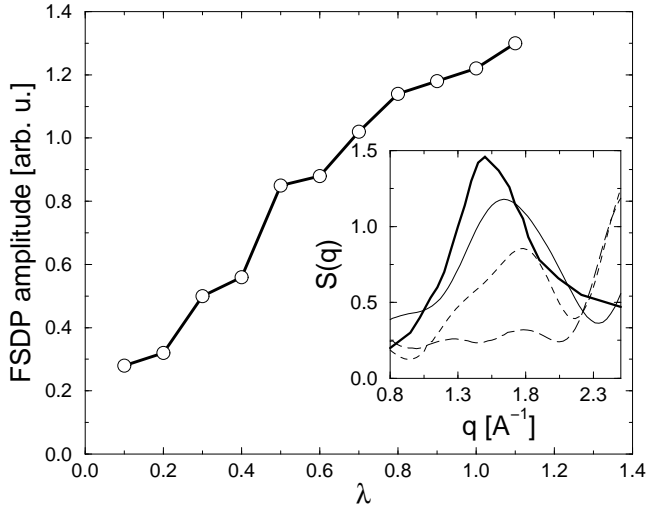,width=8.5cm,height=7cm}
\caption{First Sharp Diffraction Peak amplitude as a function of $\lambda$.
In inset the structure factor S(q) is shown in the vicinity of the FSDP for
$\lambda$=0.2 (long dashed line), $\lambda$=0.5 (dashed line),  
$\lambda$=1.0 (thin solid line) and compared to the experimental S(q) (bold 
solid line) for SiO$_2$.
}
\label{Fig3}
\end{figure}

\vspace*{1cm}
\begin{figure}
\psfig{file=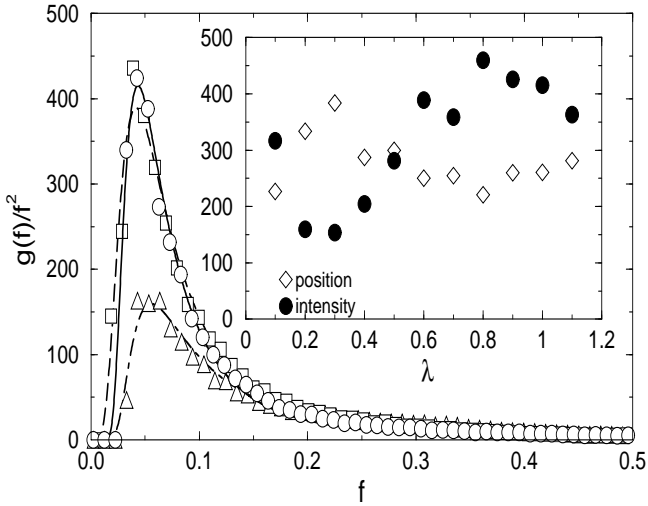,width=8.5cm,height=7cm}
\caption{ The Boson peak represented as
a plot of $g(f)/f^2$ versus $f$, where $g(f)$ is the normalized density of vibrational states and $f$ the ``reduced frequency'', taken equal to $\nu/\nu_{\rm max}$, where $\nu_{\rm max}$ is the maximum frequency of the spectrum.
$\triangle$, $\Box$ and {\Large $\circ$} correspond to $\lambda =$ 0.2, 0.6 and 1.0 respectively.
Dot-dashed, dashed and continuous lines are the corresponding fits using formula (3).
In inset the position ($\Diamond$) and the intensity ({\Large $\bullet$}) of the Boson peak are shown as a function of $\lambda$.
}
\label{Fig4}
\end{figure}


\end{document}